\documentclass[]{mn2e}
\usepackage{times}
\usepackage{graphicx,subfigure,epsf}

\def\xmm{{\it XMM-Newton\/}}
\def\cha{{\it Chandra\/}}
\def\etal{et al.\ }
\def\betamod{$\beta$-model}

\def \h50 {$h_{50}$}
\def \h70 {$h_{70}$}
\def \exo {EXO~0423.4$-$0840}
\def \Cha {{\it Chandra\/}}
\def \ciao {{\sc ciao}}

\def \nh {$N_{\rm H}$}
\def \mekal {{\sc mekal}}

\begin{document}
\title[On the nature of \exo]{On the nature of \exo}
\author[E. Belsole \etal]{E. Belsole\thanks{E-mail:
e.belsole@bristol.ac.uk}, M. Birkinshaw,  D. M. Worrall\\
Department of Physics, University of Bristol, Tyndall Avenue,
Bristol BS8 1TL, UK\\
}

\date{Accepted . Received ; in original form }
\maketitle

\begin{abstract}
We present a \Cha\ observation of the candidate BL Lac object \exo. The X-ray emission from \exo\ is clearly extended, and  is associated with an optical early-type galaxy, MCG-01-12-005, at the centre of  cluster ClG 0422-09. We do not detect a point source which can be associated with a BL Lac, but we found a small radio source in the centre of MCG-01-12-005.
The cluster gas temperature mapped by the \Cha\ observation drops continuously
from 80 kpc towards the centre, and is locally single phase. We measure a  metallicity profile which declines outwards with a  value 0.8$Z_{\odot}$ in the centre, dropping to 0.35$Z_{\odot}$ at larger radius, which  we interpret as a superposition of cluster gas and a dense interstellar medium (ISM) in the central galaxy. Although the temperature profile suggests that conduction is not efficient, the ISM and intra-cluster medium seem not to have mixed.
The entropy profile declines continuously towards the centre, in agreement with recent results on groups and clusters.
The radio source appears to have had some effect in terms of gas heating, as seen in the small scale ($\sim$10 kpc) entropy core, and the asymmetric hard emission on the same scale.
\end{abstract}

\begin{keywords}
galaxies: individual: MCG-01-12-005 and EXO~0423.4$-$0840 --- X-ray: galaxies: interstellar medium --- radio: galaxies --- galaxies: active: BL Lacertae --- galaxies:clusters:intergalactic medium:individual: EXO~0422, ClG0422-09 
\end{keywords}

\section{Introduction}\label{intro}
\exo\  was discovered as part of an EXOSAT high galactic latitude survey (Giommi et al. 1991). The source was associated with the $m_{pg}$ = 15.9 mag galaxy  MCG-01-12-005, a radio source (with 6 cm flux density 44 mJy)  with no optical emission lines. It was classified as a candidate BL Lacertae object on the basis of its X-ray luminosity (above $10^{43}$ erg s$^{-1}$) and its position in the $\alpha_{0x}-\alpha_{r0}$ diagram, and is included as a BL Lac object in the V\'eron-Cetty \& V\'eron (2003) catalogue.  Observations with CANGAROO give an upper limit for the $\gamma$-ray flux above 2 TeV of less than $4\times10^{-12}$  photons cm$^{-2}$ s$^{-1}$ (Roberts et al. 1998) but the results are not conclusive on the nature of the source. 
Kirhakos \& Steiner (1990) determined a redshift of 0.0392 for the source they designate IRAS 04235-0840B and identify with the HEAO source 1H 0422-086, which they classify as a type 2 Seyfert, and associate with \exo. 

MCG-01-12-005 is an early type galaxy close to the centre of galaxy cluster ClG\,0422-09, at $z=0.039$. White, Jones \& Forman (1997) described it as a cooling-flow cluster and give a temperature of 2.9 keV (from Edge \& Stewart 1991). ASCA data analysed by Ikebe et al. (2002) implied  a temperature of k$T=2.0^{+0.9}_{-0.6}$ keV from the $L_X-T$ relation, and a 0.1-2.4 keV luminosity of (5.3$\pm0.5)\times10^{43}$ erg s$^{-1}$. The Galactic absorption for \exo\ is 6.22$\times10^{20}$ cm$^{-2}$ (Dickey \& Lockman 1990), and the $N_{\rm H}$ reported in X-ray analysis of the galaxy cluster is 6.4$\times10^{20}$ cm$^{-2}$ (e.g Ikebe et al.  2002).

In this paper we present \Cha\ observations of \exo\, which give us new insights into the nature of this source.
Throughout this paper, we adopt a redshift for \exo\ of $z=0.039$ and a cosmology with $H_0$ = 70 km s$^{-1}$ Mpc$^{-1}$, $\Omega_m=0.3$ and $\Omega_{\Lambda}$ =0.7. The angular diameter distance to this source is 159 Mpc, and its luminosity distance is 172 Mpc. At the redshift of the source, 1 arcsec corresponds to 773 pc.

\section{Observations and preparation}
\begin{figure*}
\centering
\subfigure[] 
{
    \label{fig:imaflux}
  \includegraphics[scale=0.47,angle=0,keepaspectratio]{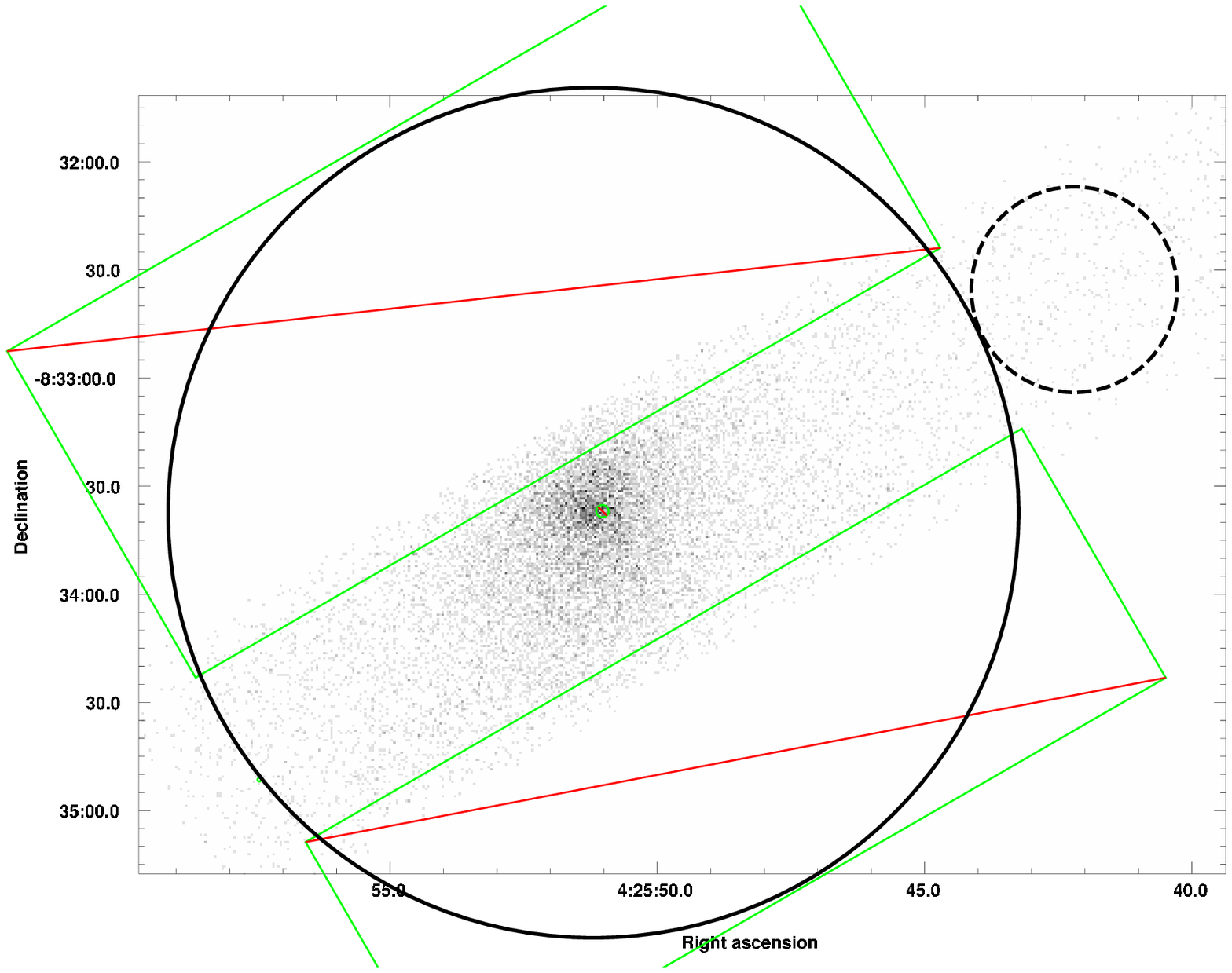}
}
\hspace{0cm}
\subfigure[] 
{
    \label{fig:dss}
    \includegraphics[scale=0.44,angle=0,keepaspectratio]{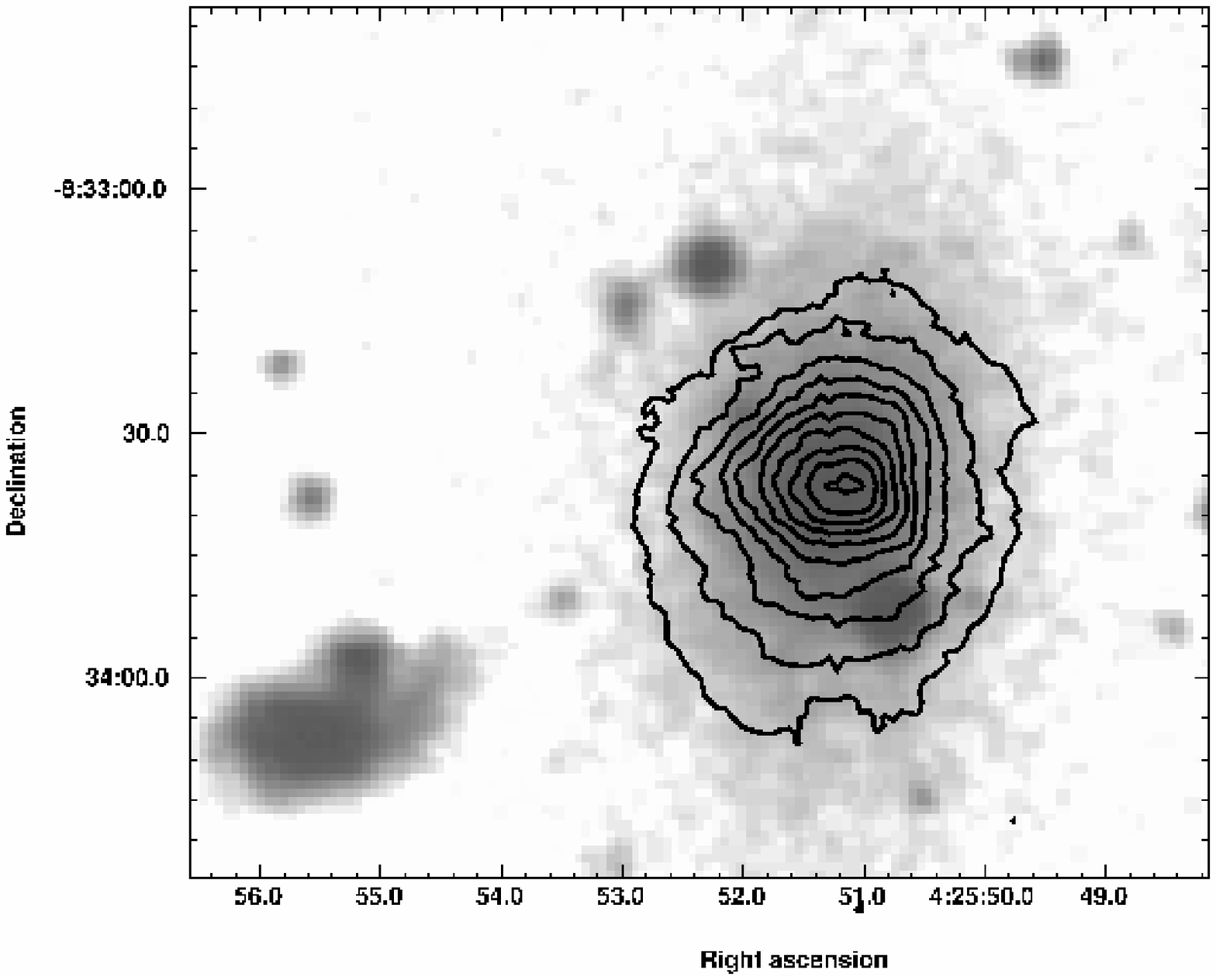}
}
\caption{{\bf (a)}:  ACIS  image of \exo\ in the energy range 0.4-7.0 keV. The solid and dashed circles indicate the regions used to extract the global and background spectra (see text). The rectangular regions were excluded to avoid the CCD boundaries. The point source detected by {\sc wavdetect} is also excluded. {\bf (b)} X-ray contours from the 0.4-7.0 keV flux image of \exo\ overlaid on the Digitized Sky Survey (DSS) optical  image.  X-ray contours are drawn at multiples of 1.1$\times10^{-7}$ photons cm$^{-2}$ s$^{-1}$ pixel$^{-1}$. The contours were obtained after excising the compact source and the extended structure $A$ (Fig. \ref{fig:imagauss}), and by replacing them with Poisson samples from the local surrounding (see text for details). We  note the different X-ray and optical elongations of the central region. The  contours are flattened to the north-west because  of the edge of the ACIS/CCD. The galaxy MCG-01-12-006, is visible to the south-east.}
\label{fig:images} 
\end{figure*}

\subsection{{\bf {\em Chandra}} data}
\exo\  was observed with \Cha\  in December 2002 (Observation ID 03970) as part of a small sample of BL Lac objects with relatively weak cores, which included PKS 0521$-$356 (Birkinshaw, Worrall, Hardcastle 2002), with the primary objective of studying the interaction between the jet and the group or cluster X-ray medium.  Since the BL Lac core of \exo\ was expected to produce a high X-ray count rate, the observation was performed in faint mode and with a small window with the back-illuminated chip S3 on the Advanced CCD Imaging Spectrometer (ACIS). This window was 128 pixels (of 0.492$\times$0.492 arcsec$^2$ each) wide, and 1024 pixels long.  We processed the level 1 event data using  \ciao\ v3.0.2 and calibration database (CALDB) 2.26. We removed bad pixels using the supplied bad pixel file, and retained events of grade 0,2,3,4, and 6. We verified that there are no periods of statistically higher background by generating light curves in different energy bands, and the useful exposure time is 14280 s. The online aspect calculator\footnote{\tt http://cxc.harvard.edu/ciao/threads/arcsec\_correc\\tion/} shows  that no aspect correction was necessary for these data, therefore the X-ray, radio and optical frames should align to the \cha\ aspect uncertainty $\sim0.6$ arcsec.

\subsubsection{Spatial analysis}
The analysis of the diffuse emission started with our use of {\sc wavdetect} to identify possible point sources in the field. The algorithm was used with a source detection threshold\footnote{For detail see \tt http://cxc.harvard.edu/ciao/download/doc/\\detect\_html\_manual/vtp\_ref.html\#par:wav\_sigthresh} set to 10$^{-9}$ (the minimum value for the algorithm to give an accurate enough detection), and only found a single point source at RA = $ 04^h25^m51^s04$, Dec = $-08^{\circ}33\arcmin36\farcs89$ (J2000).

The characterisation of the diffuse emission is affected by the half-chip observation, which induces border effects to the north-east and south-west. We corrected these effects via the exposure map, generated using the aspect histogram and the instrument map. When appropriate, the counts image was converted into a flux image by dividing it by the exposure map. 

\subsubsection{Spectral analysis}
For spectral analysis, we accounted for the position-dependent redistribution matrix file (RMF) and the auxiliary response file (ARF) for the source, which were computed from \Cha\ calibration database 2.26. This release takes into account a correction for the degradation of the low-energy quantum efficiency. We extracted spectra in successive circular annuli, excluding the point source and small scale structures (see Sec. \ref{sec:centre}). The background for all annuli was chosen to be the same, and was extracted in a circle of radius 28.5 arcsec (see Fig. \ref{fig:imaflux}) centred at RA = $04^h25^m42^s268$, Dec = $-08^{\circ}32\arcmin35\farcs91$ (J2000). For the spectral analysis we excluded the regions close to the chip boundary. If not otherwise specified, spectra were fitted with absorbed plasma models in XSPEC v11.3. We tried fits using the APEC (Smith et al. 2001) and \mekal\ (Mewe, Gronenschild, van den Oord 1985; Liedahl, Osterheld, Goldstein 1995) codes; the latter gave better fits. We scaled abundances to the standard solar values from Grevesse \& Sauval (1998).

\begin{figure*}
\begin{centering}
    \includegraphics[scale=0.85,angle=0,keepaspectratio]{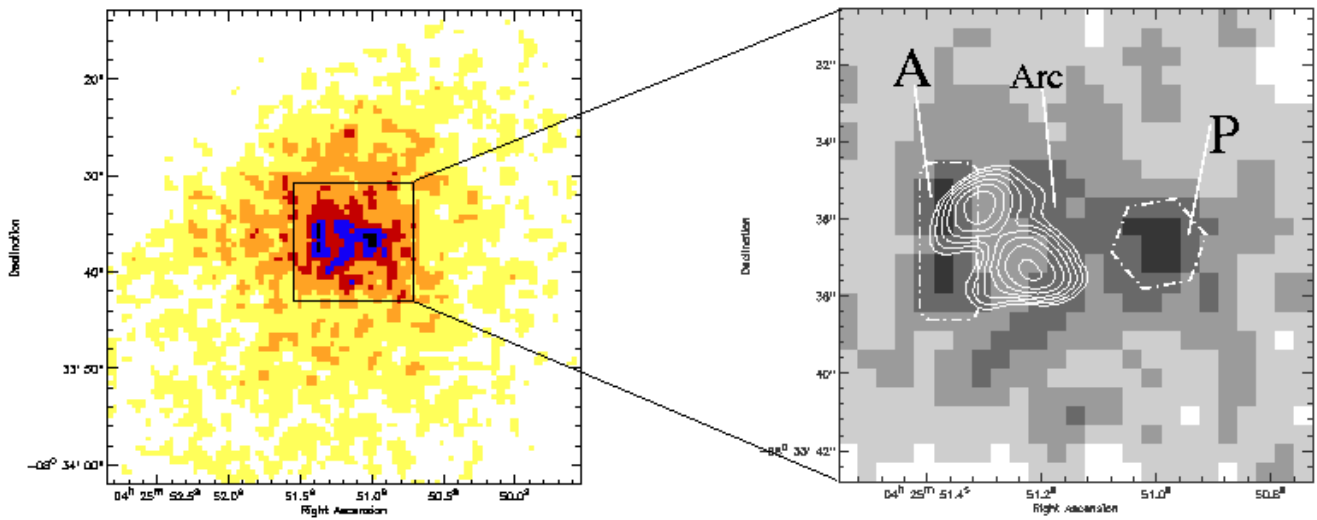}
\caption{Image in the energy range 0.4-7.0 keV smoothed by a Gaussian filter of $\sigma=2$ pixels. The insert shows a zoom of the same image with overlaid  contours of the  4.9 GHz map. The first contour is traced at 0.5 mJy, the last at 10 mJy,  and they are logarithmically spaced. The figure shows the complex X-ray structure present in the core of the cluster. The radio emission lies in a  central deficit in the X-ray surface brightness. $P$ indicates the compact X-ray source detected by {\sc wavdetect}. The $Arc$ is marginally significant, while the area labelled with $A$ will be discussed in Sec. \ref{sec:centre}. }
\label{fig:imagauss}
\end{centering}
\end{figure*}

\subsection{Radio data}
A-array VLA~data for EXO~0423.4$-$0840 at 1.4, 4.9, and~8.6~GHz were extracted from the National Radio Astronomy Observatory (NRAO) archive, and processed into maps through the normal processes of calibration and CLEANing. At 1.4 and 4.9~GHz
it was possible to self-calibrate the data to improve the final maps, but the high noise and short observation at 8.6~GHz made this impossible.

All three maps easily detect a bright extended radio source near the centre of the X-ray emission. At 1.4~GHz the radio map has a synthesized beam of $2.3 \times 1.9$~arcsec, and does not well define the structure of the radio source. However, at this lowest frequency we also find a diffuse source located near RA = $\rm 04^h25^m55^s$, Dec =$-08^\circ34^\prime 05^{\prime\prime}$ which may (or may not) be associated with \exo. Unfortunately, this structure lies off the field of the {\it Chandra} X-ray observation.

The 4.9-GHz image of \exo\  (Fig.~2) shows the radio emission to be from a small double source. At this resolution ($0.9 \times 0.5$~arcsec) the two components are well separated, but it is unclear based on the radio data whether one of the components is a  core and the other is a one-sided jet or both components are lobes symmetrically disposed about the  core.

The 8.4-GHz image has $0.4 \times 0.3$~arcsec full-width-half-maximum (FWHM) but a high noise level, so that we cannot currently distinguish which component has the flatter spectrum, or whether there is a core present between the two components. Further high-resolution, higher-sensitivity radio data are needed to resolve this issue.

\section{X-ray morphology}\label{sec:morphology}

\exo\ shows clear diffuse X-ray emission (see Fig. \ref{fig:imaflux}). The X-ray  peak is at RA = $04^h25^m51^s0$, Dec = $-08^{\circ}33\arcmin36\farcs9$ (J2000) which is close (less than 5 arcsec) to the NASA/IPAC Extragalactic  Database (NED) position of  galaxy MCG-01-12-005, and coincident with the point source detected by {\sc wavdetect}\footnote{At the time of writing  NED  associates \exo\ with  galaxy pair MCG-01-12-006, and gives coordinates for \exo\ of  RA = $04^h25^m55^s56$, Dec = $-08^{\circ}34\arcmin07\farcs5$ (J2000). The \cha\ data show this to be an error, since no X-ray emission is associated with MCG-01-12-006. SIMBAD correctly associates \exo\ with galaxy MCG-01-12-005.}. The centre of the cluster of galaxies ClG 0422-09 is at J2000 coordinates RA = $04^h25^m51^s02$, Dec = $-08^{\circ}33\arcmin38\farcs5$, less than 2 arcsec from the \Cha\ X-ray peak. The \Cha\ image suggests that the bulk of the X-ray emission from \exo\ is not due to the expected  BL Lac object, but is diffuse in nature and originates from the intergalactic medium (IGM)  of  galaxy MCG-01-12-005, and intracluster medium (ICM) of galaxy cluster ClG\,0422-09.

The optical emission of MCG-01-12-005 from the Digital Sky Survey (DSS; Fig. \ref{fig:dss}) is elongated in the N-S direction, while the X-ray emission in the central regions displays a slight elongation in the E-W direction. The X-ray contours become more circular at larger distance from the centre. 

A more extended structure lies east of the X-ray peak and causes the elongation of the central region of the X-ray contours. We applied a Gaussian filter of FWHM 0.98 arcsec to the (0.4-7.0 keV) image, to enhance the extended structure (see Fig. \ref{fig:imagauss}). Two X-ray enhancements to  the east of the X-ray peak, labelled with $Arc$, seem to be connected to the compact source (labelled with $P$). However this structure  is barely  significant. The other extended structure (marked as $A$ in the insert of Fig. \ref{fig:imagauss}) is significant at 11$\sigma$ above the surrounding area. In this Figure we also show the radio contours from the 4.9 GHz map. We observe that the radio emission is located in a region of lower X-ray  surface brightness than its immediate surroundings.  We note that the optical centre of the galaxy  and the southern peak of the radio emission at 4.9 GHz coincide, suggesting that this is a likely location for the X-ray centre of the object. Unfortunately, the radio data are not sufficient  to define  the location of the core emission.

To avoid difficulties arising from the uncertain location of the X-ray  centre, we excluded structure $A$ and the compact source ($P$) for our analysis of the large scale structure. We will discuss the structure at the centre in a separate section. We replaced these two structures with their local surrounding mean counts and calculated the centroid of the resulting X-ray emission, which was found to be at  RA = $04^h25^m51^s26$, Dec = $-08^{\circ}33\arcmin36\farcs9$ (J2000). For any extended source analysis in the following, this will be fixed as the centre  of the \Cha\ data.

\section{Cluster emission}
\subsection{Surface-brightness profile}\label{sec:sbprof}
\begin{figure}
\begin{centering}
    \includegraphics[scale=0.75,angle=0,keepaspectratio]{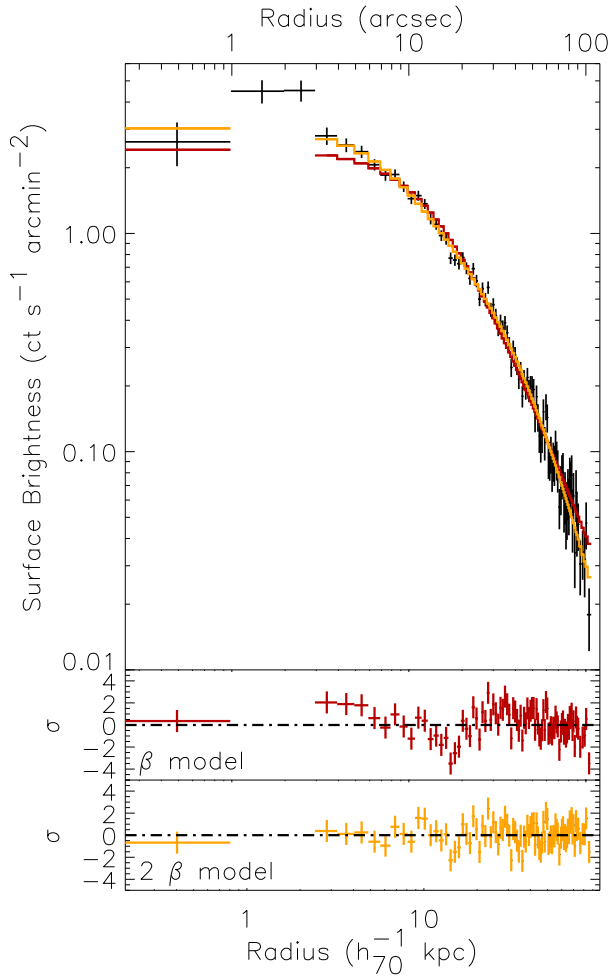}
\caption{Surface brightness profile derived by integrating counts in the 0.4-2.0 keV energy range. The profile was binned to have a minimum S/N=3 in each bin. The bottom panels show the residual from the 1-$\beta$ and the 2-$\beta$ models. Bins 2 and 3 were excluded when fitting.}
\label{fig:sbprof}
\end{centering}
\end{figure}

We extracted a surface-brightness profile  by integrating photons in concentric 1-arcsec wide annuli (masked by the rectangular regions in \ref{fig:imaflux}) out to  118 arcsec and in the 0.4-2.0 keV energy range. We subtracted a local background represented by the (masked) annular region between 120 and 130 arcsec. The inner 3  arcsec of the profile display some fine-scale structure (see Fig. \ref{fig:sbprof}), presumably from errors in the interpolation over structures  $P$ and $A$ as discussed above. The excess emission represented by bins 2 and 3 has little effect on the parameters of fits to the overall structure of the cluster, but it increases the total $\chi^2$. We thus  excluded these two bins when fitting. We binned the surface-brightness profile to have a signal-to-noise (S/N) of at least 3$\sigma$ above the background in each bin. We then initially fitted the profile with a 1D \betamod:
\begin{equation}
S(r) = S_0 \left [1+\left(\frac{r}{r_c}\right)^2\right ]^{-3\beta+1/2}
\label{eq:beta}
\end{equation}
where $r_{\rm c}$ and $\beta$ are the core radius and the index, and $S_{\rm 0}$ is the central surface brightness.  The model in equation \ref{eq:beta} is not a good representation of the data, providing  best-fitting parameters of $S_{\rm 0} = 2.4\pm0.1$ counts s$^{-1}$ arcmin$^{-2}$, $\beta = 0.51\pm0.01$, $r_{\rm c} = 14.0\pm0.8$ arcsec ([10.7$\pm0.6$]$h_{70}^{-1}$ kpc) for a reduced $\chi^2$, $\chi_{\nu}^2$ =  1.64 = 147.9/90 d.o.f. (the $\chi_{\nu}^2$ rises to 1.95 if bins 2 and 3 are taken into account, but $\beta$ and $r_{\rm c}$ do not change within the errors).

The fit improved when we fitted the profile with a sum of two $\beta$ models, 
\begin{equation}\label{eq:2beta}
\begin{array}{lcl}
S(r)&=& S_1(r)+S_2(r).\\
\end{array}
\end{equation}
where $S_1$ and $S_2$ are defined as in Eq. \ref{eq:beta}.
We required the $\beta$ parameters of both models  to be the same in order to have a smooth power law shape for the density profile\footnote{We also tried fits with two separate $\beta$ parameters, but the $F$-test indicates no significant improvement in the fits when the second $\beta$ is allowed to vary independently from the first.}  at large radii. 
This model gives an excellent fit of $\chi_{\nu}^2$ = 1.03 =  90.2/88 d.o.f., with  $S_{01} =  2.25\pm0.15$ counts s$^{-1}$ arcmin $^{-2}$, $r_{\rm c1} = 10.8\pm1.8$ arcsec ([8.3$\pm1.4$] $h_{70}^{-1}$ kpc), $S_{02} =  0.79\pm0.11$ counts s$^{-1}$ arcmin $^{-2}$, $r_{\rm c2} = 38.4\pm6.0$ arcsec ([29.7$\pm4.6$]$h_{70}^{-1}$ kpc), and $\beta=0.71\pm0.06$. For comparison, the analysis of {\it ROSAT} data by Reiprich \& B\"ohringer (2002) found a value of $\beta = 0.722^{+0.104}_{-0.071}$ and $r_{\rm c} = 101^{+29}_{-21}$ \h70 $^{-1}$ kpc.  The $\beta$ values are in good agreement if the two-$\beta$ model is considered. In Fig. \ref{fig:sbprof} we show the data points and both the single $\beta$ and double $\beta$ best-fit models. The lower two  windows show the residuals. We derive the gas density distribution in combination with the spectral analysis.

\subsection{Global spectrum}
We extracted the global spectrum of the source by selecting a circular region of radius 118 arcsec (see Fig. \ref{fig:imaflux}). To limit spurious effects produced by the boundary of the exposed part of the chip, we excluded two rectangles of size  $260\times105$ arcsec$^2$ centred on RA = $04^h25^m53^s9$, Dec = $-08^{\circ}32\arcmin41\farcs8$, and RA = $04^h25^m48^s5$, Dec = $-08^{\circ}34\farcm49\farcs5$ (J2000), respectively with the same orientation as the CCD (position angle = 30$^{\circ}$). The compact source was also excluded. We took  background from the usual circle (Sec. 2.1.2; Fig. \ref{fig:imaflux}).

\begin{figure}
\begin{centering}
\includegraphics[scale=0.37,angle=-90,keepaspectratio]{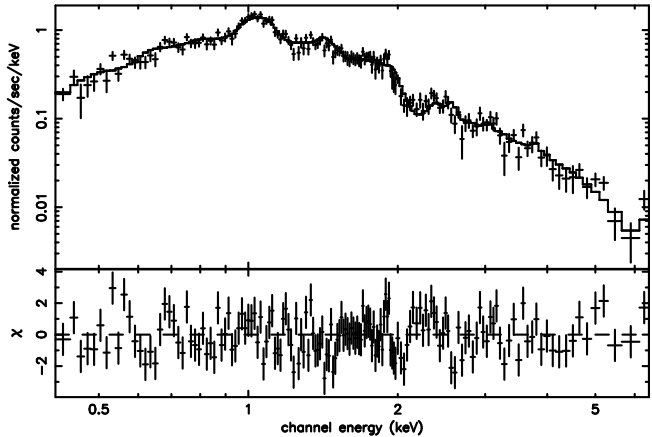}
\caption{Background subtracted global spectrum of \exo\, extracted in a circle of 118 arcsec. The folded model is a two-temperature \mekal\ with absorption as a free parameter. The 1.9-2.1 energy bin was ignored when fitting due to uncertainties in the Ir instrumental features.}
\label{fig:spglob}
\end{centering}
\end{figure}

The source has a 0.4-7.0 keV count rate of 1.27 cts s$^{-1}$ in this region. We binned the spectrum  to  60 counts per bin prior to background subtraction. The global spectrum (see Fig. \ref{fig:spglob}) shows strong emission at the energy corresponding to the Fe L complex. Other spectral features are also detected, although some of them are instrumental (e.g. the region 1.9-2.1 keV near the Ir edge) and are excluded when fitting.
We fitted the global spectrum between 0.4 and 7.0 keV with thermal models in XSPEC (version 11.3.0), leaving  the fractional abundance relative to solar free to vary, and initially adopting a Galactic absorption of $N_{\rm H} = 6.2\times10^{20}$ cm$^{-2}$ (Dickey \& Lockman 1990). 

A single \mekal\ model gives a bad fit with $\chi^2 = 322.0/148$ d.o.f., a temperature of about 2.8 keV and a metallicity of 1.4 solar, with large residuals at low energy. A much improved, though still unacceptable fit, with $\chi^2 = 234/147$ d.o.f., is obtained if the absorption is allowed to vary. The resulting \nh = (1.19$_{-0.10}^{+0.17})\times10^{21}$ cm $^{-2}$ is roughly double the expected Galactic absorption.

An alternative explanation for the poor fit at low energy might be thought to be systematic errors in the used (version 2.26) calibration of the quantum efficiency degradation of the ACIS-S3 chip. To test this we repeated the fit but used only data above 1 keV. The resulting fit (Table 1, column 3) is statistically indistinguishable from a fit using the full energy range, and so we concluded that the high \nh\ is not an artifact of the data processing.

Fitting a two-\mekal\ model with  \nh\ as a free parameter in the 0.4-7.0 keV range gives $\chi^2=211.3/145$ d.o.f. The $F$-test indicates that the two-temperature model is  a better representation of the data at 99.9\% significance (see Fig. \ref{fig:spglob}). The best-fitting results are listed in Table \ref{tab:spglobfit}.  The 0.2-10.0 keV total unabsorbed luminosity is 3.7$\times10^{43}$ erg s$^{-1}$. A model composed of a \mekal\ plus a power law gives a better fit  ($\chi^2$/d.o.f.=222.2/145) than the single temperature representation, but does not improve on the two-temperature model (see Table \ref{tab:spglobfit}) and the spectral index is not well determined. We thus do not discuss this model any further.

\begin{table*}
\begin{center}
\caption{Best-fitting results of the global spectral analysis.}
\label{tab:spglobfit}
\begin{minipage}{400pt}
\begin{tabular}{l|cccccc}
\hline
Parameter & 1T  & 1T   & 1T  & 2T & 1T+1PL & 1T+1CF\\
& (0.4-7.0 keV) & (0.4-7.0 keV) & (1.0-7.0 keV)& (0.4-7.0 keV)& (0.4-7.0 keV)& (0.4-7.0 keV) \\
\hline
\nh\ ($10^{21}$ cm$^{-2})$& 0.62 (fixed) & 1.19$^{+0.17}_{-0.10}$ & 1.20$^{+0.60}_{-0.58}$  &1.42$^{+0.27}_{-0.18}$ &1.27$^{+0.49}_{-0.13}$& 1.27$^{+0.13}_{-0.12}$\\
kT1 (keV)& 2.83 &2.36$^{+0.14}_{-0.20}$	&  2.50$^{+0.22}_{-0.25}$ &2.47$^{+0.16}_{-0.17}$ & 2.00$^{+0.19}_{-0.19}$ &2.61$^{+0.19}_{-0.19}$\\
$Z/Z_{\odot}$ & 1.42 & 0.72$^{+0.15}_{-0.20}$ & 0.90$^{+0.24}_{-0.20}$ &0.91$^{+0.22}_{-0.17}$ & 0.60$^{+0.90}_{-0.18}$&0.88$^{+0.23}_{-0.18}$\\
$N^*$ ~($\times10^{-3}$)& 6.12 & 8.49$^{+1.03}_{-0.59}$&   7.77 $^{+0.75}_{-0.70}$& 7.80$^{+0.70}_{-0.68}$&7.72$^{+1.36}_{-6.80}$ &7.09$^{+0.69}_{-0.73}$\\ 
kT2 (keV) & --- & --- &    --- &0.38$^{+0.15}_{-0.09}$ & ---& ---\\
$N^*$ ~($\times10^{-3}$)& --- & --- & --- &0.33$^{+0.32}_{-0.16}$ & --- & ---\\
$\Gamma$  (keV) & --- & --- & --- &--- & 1.54($<2.34$)&---\\
$N$ ~($\times10^{-3}$)& --- & --- &--- &--- & 0.31$(<1.90)$&---\\
k$T_{\rm low}$ (keV) & --- & --- & --- &--- & --- &  0.08$^{+0.45}_{-0.08}$\\
$N$& --- & --- & --- &--- & --- &10.0$^{+4.0}_{-4.0}$ \\
$\chi^2$/d.o.f. & 322.0/148&234.0/147& 153.5/107 & 211.3/145 & 222.2/145 &216.3/145\\
\hline
\end{tabular}

Abundances are from Grevesse and Sauval (1998). In the two-temperature \mekal\ model the abundances of the two thermal components were tied. $N$ is the normalisation given by  XSPEC, as $^*$ $K ({\rm cm}^{-5}) = 10^{-14} / (4 \pi (D_A\times(1+z))^2) \int n_e n_H dV$, where $D_A$ is the angular size distance to the source (cm), $n_e$ is the electron density (cm$^{-3}$), and $n_H$ is the hydrogen density (cm$^{-3}$). Errors are quoted at 90\% confidence for one interesting parameter. The energy range used for fitting is listed for each fit.
\end{minipage}
\end{center}
\end{table*}

The results on the global spectrum suggest the following:
\begin{enumerate}
\item  the gas spectrum is consistent with a two-temperature model. Spectral fitting to ASCA data (see e.g.  Ikebe et al. 1999, Makishima et al. 2001) suggested that a two-temperature model can mimic a multi-temperature gas as well as a cooling flow (CF), and the contribution of the cD galaxy has to be taken into account. We thus fitted the global spectrum with a model composed of a \mekal\ plus {\sc mkcflow} (Mushotzky \& Szymkowiak 1988). The high temperature of the CF model was fixed to be the same as the \mekal\ model temperature. The metallicities were forced  to be the same and the absorbing  column density left free.
The best-fit has $\chi^2=216.3/145$ d.o.f. The high temperature is k$T=(2.61\pm0.19)$ keV, the low temperature is poorly constrained, with a best-fit value of k$T = 0.08$ keV (and upper limit of 0.45 keV), the metallicity $Z/Z_{\odot}=0.88^{+0.23}_{-0.18}$, and the \nh = (1.27$^{+0.13}_{-0.12})\times10^{21}$ cm$^{-2}$, all  in agreement, at 90\% confidence, with the two-temperature model described above, though a worse fit. The mass deposition rate is 10$\pm4$ M$_{\odot}$ yr$^{-1}$.

\item the absorption is higher than Galactic by more than a factor of 2. As shown later (Sec. 4.3.1) the excess absorption encompasses the whole field of view and not only the cool central part of the cluster, suggesting that this excess absorption is of Galactic rather than intrinsic origin. No CO map in the direction of our source is good enough to establish if molecular gas contributes the needed column. However, there is a good correlation between  Galactic HI and the infra-red (IR)  emitting dust (see Boulanger et al. 1996). Pointecouteau et al. (2004) found a remarkable agreement between their X-ray fitted \nh\ profile, and the total $N_{\rm H}$ profile derived from this  correlation.
To test if this interpretation is applicable to \exo, we looked at the Infra-Red Astronomical Satellite (IRAS) map at 100 $\mu$m, in the direction of the source (Schlegel,  Finkbeiner, \&  Davis 1998). The map shows bright clumpy emission, with strong gradients. 
 The IRAS map is at far larger spatial scales than the \Cha\ field of view, but the  100 $\mu$m flux in the vicinity of the field is $\sim 7.2$ MJy sr$^{-1}$. The IR - \nh\ relation of Boulanger et al. (1996) then  predicts  $N_{\rm H}\sim1.4\times10^{21}$ HI cm$^{-2}$. This is in good agreement with our best-fit results from the global \Cha\ spectrum, suggesting the absorption excess is, indeed, from molecular gas.
\end{enumerate}

\subsection{Temperature and abundance profiles}\label{sec:tprof}
In order to perform spatially resolved spectroscopy, we selected concentric annular regions out to 118 arcsec, masked by the rectangles seen in Fig. \ref{fig:imaflux}. The first bin is  of radius 5 arcsec, just outside the bright structures in the centre. Possible variations of the spectral modelling  within this region  are discussed in Section \ref{sec:centre}, but they do not affect our conclusions here.
\begin{figure}
\begin{centering}
   \includegraphics[scale=0.5,angle=0]{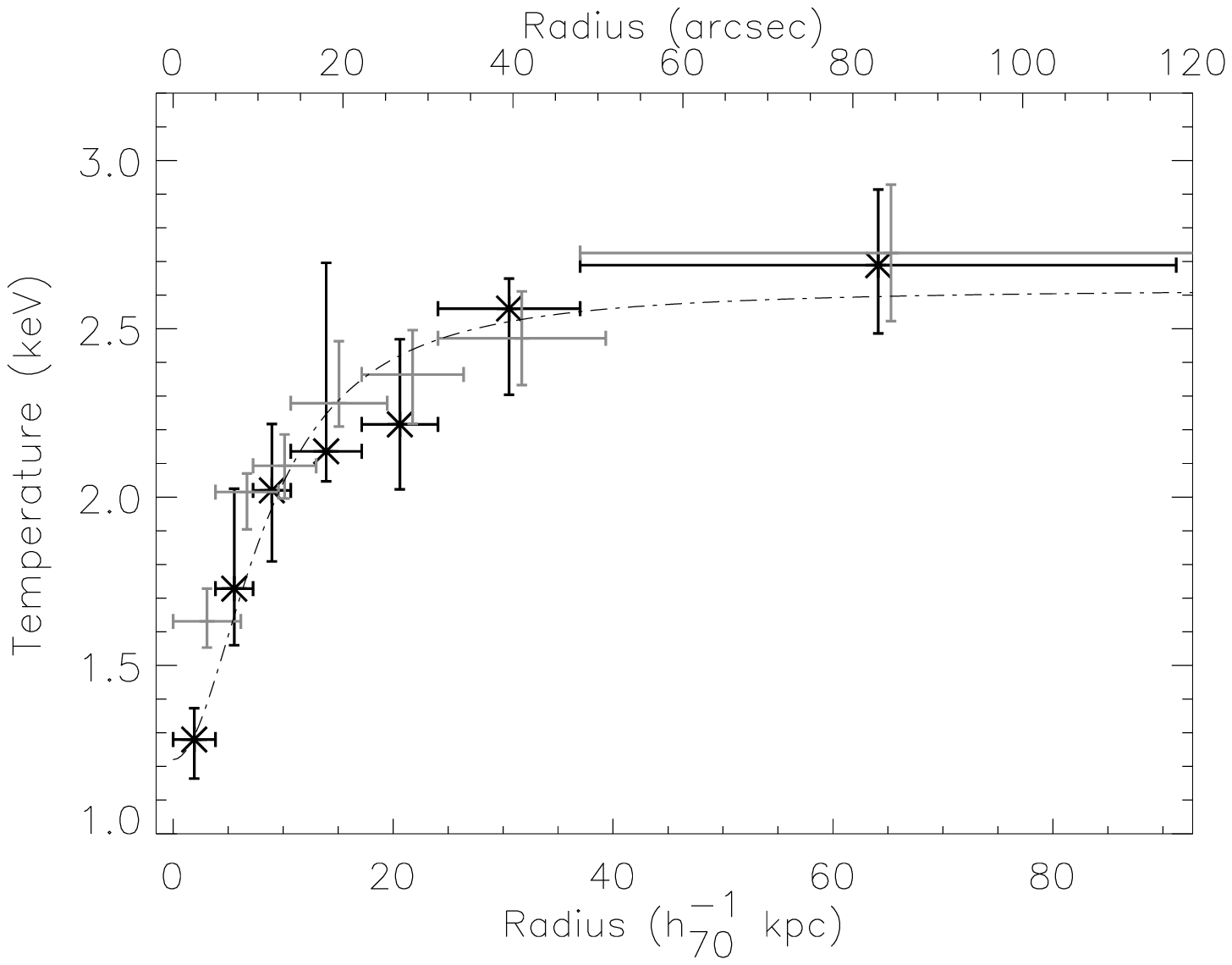}
    \includegraphics[scale=0.5,angle=0]{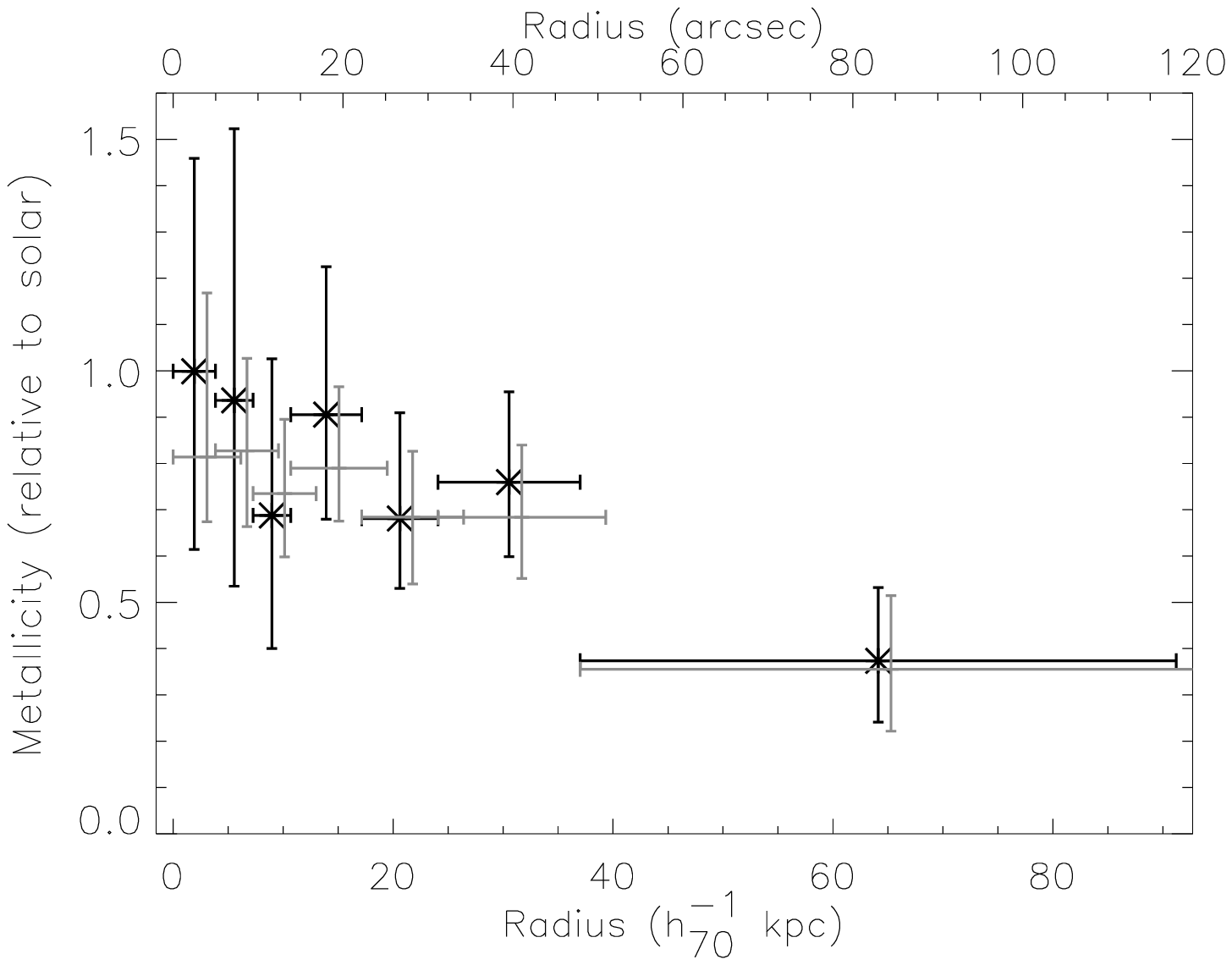}
\caption{ {\bf Top:} Projected (grey) and deprojected  (black) temperature profiles of \exo. The two data sets are shifted  for better visibility. The dash-dotted line is the best fit of Equation \ref{eq:tproffit} to the deprojected data points (see Sec. \ref{sec:PS}); {\bf Bottom}: Projected (grey) and deprojected  (black) metallicity  profiles.}
\label{fig:tzprof}
\end{centering}
\end{figure}

\subsubsection{Projected quantities}\label{sec:tproj}
We obtained a projected temperature profile by fitting each annulus with a \mekal\ model. We used the same background  as for the global spectrum. All spectra were binned to 15 counts per bin prior to background subtraction, and fitted between 0.4 and 7.0 keV. 

In order to test if \nh\ is constant within the area of the X-ray detection, we fitted each annulus with a \mekal\ model with \nh, temperature and abundances as free parameters. We found that all the column densities  are consistent with the value found in the global analysis (within 1$\sigma$ errors).
To strengthen this result, we fitted the spectrum in each annulus by  excluding data below  1 keV. (This also tests for  possible spatial variation of the quantum efficiency degradation at low energy.) These fits are consistent with those in the full energy range. We thus confirm that \nh\ is a factor of two higher than the value predicted by the HI measurements at all radii, and that it is constant over a scale $\sim120$ arcsec (90 kpc), with a global value of $N_{\rm H} = (1.19^{+0.17}_{-0.10})\times 10^{21}$ cm$^{-2}$. In the following the column density is fixed to this value.

We next  fitted the spectra again in each annulus, between 0.4 and 7.0 keV,  with an absorbed \mekal\ model with \nh\ fixed. A single-temperature model is an acceptable representation of the data in each annulus. However, to test possible contamination from a non-thermal source in the inner bin, we added a second component, represented by a power law, when fitting the first bin. The additional second component is not required (we also added a second thermal model, and the conclusion is similar). 

The projected  radial temperature and metallicity profiles are shown in Fig. \ref{fig:tzprof} by the grey data points (crosses). Displayed errors are 1$\sigma$ for one interesting parameter. There is a clear temperature drop towards the centre, with a minimum projected k$T=1.7$ keV. The temperature rises almost  monotonically with radius to reach a maximum of 2.8 keV in the outermost bin, which is in agreement with the global temperature found with ROSAT (White et al. 1997). This suggests that the \Cha\ data are giving us a fine-scale measurement of the cool gas in the core region.

The projected metallicity profile drops continuously from a value of 0.8 $Z_{\odot}$ in the centre to $Z/Z_{\odot}$=0.35 in the outermost region.

\subsubsection{Correcting for projection effects}\label{sec:tdeproj}

We accounted for projection effects in the profiles of Fig. \ref{fig:tzprof} by using the {\sc projct} model in XSPEC, and assuming spherical symmetry. The volumes of each shell are calculated by the model once the inner and outer boundaries of each annulus are added to the FITS headers of each spectrum. Since the cluster emission extends well beyond our outermost annulus, the emitting volume associated with the outermost annulus will be too small. The effect is to underestimate the brightness of the next annulus in.

We fitted an absorbed,  single-temperature {\sc mekal} model in which the temperature and metallicity were left as free parameters at each radius. The column density was fixed to the global value as before. The best-fitting model gives $\chi^2$ = 802.6/791 d.o.f.

The asterisks in  Fig. \ref{fig:tzprof}  show the deprojected temperature and abundance profiles of the cluster. When projection effects are taken into account, the temperature in the central bin is significantly cooler than its respective projected value, since the hot gas from the outer regions seen in projection  is accounted for. For the outer annuli the projected and deprojected temperatures agree within the errors. 

The metallicity  declines outwards and within the errors is also consistent with a constant value of 0.8 $Z_{\odot}$ within 40 arcsec and then drops to $Z/Z_{\odot}=0.37$ in the outermost bin.

From the deprojected emission measure it is possible to calculate the ion and electron density directly from XSPEC. Under the assumption that the density is constant within each annulus, the ion density is simply:
\begin{equation}
$$n_e~n_H = N\times4\pi~10^{14}~D_A^2~(1+z)^2/V$$
\label{eq:nhfromN}
\end{equation}
where $N$ is the normalisation from the deprojected \mekal\ model fitting, $D_A$ is the angular diameter distance and $V$ is the volume of the shell.
 
The electron density is then obtained from the relation (see e.g. Worrall \& Birkinshaw 2004)

\begin{equation}
\frac{n_e}{n_H}=\frac{1+X}{2X}\sim 1.18
\end{equation}
where $X$ is the mass fraction in hydrogen, $X= 0.74$ for normal cosmic (solar) abundances.

\begin{figure}
\begin{centering}
   \includegraphics[scale=0.5,angle=0]{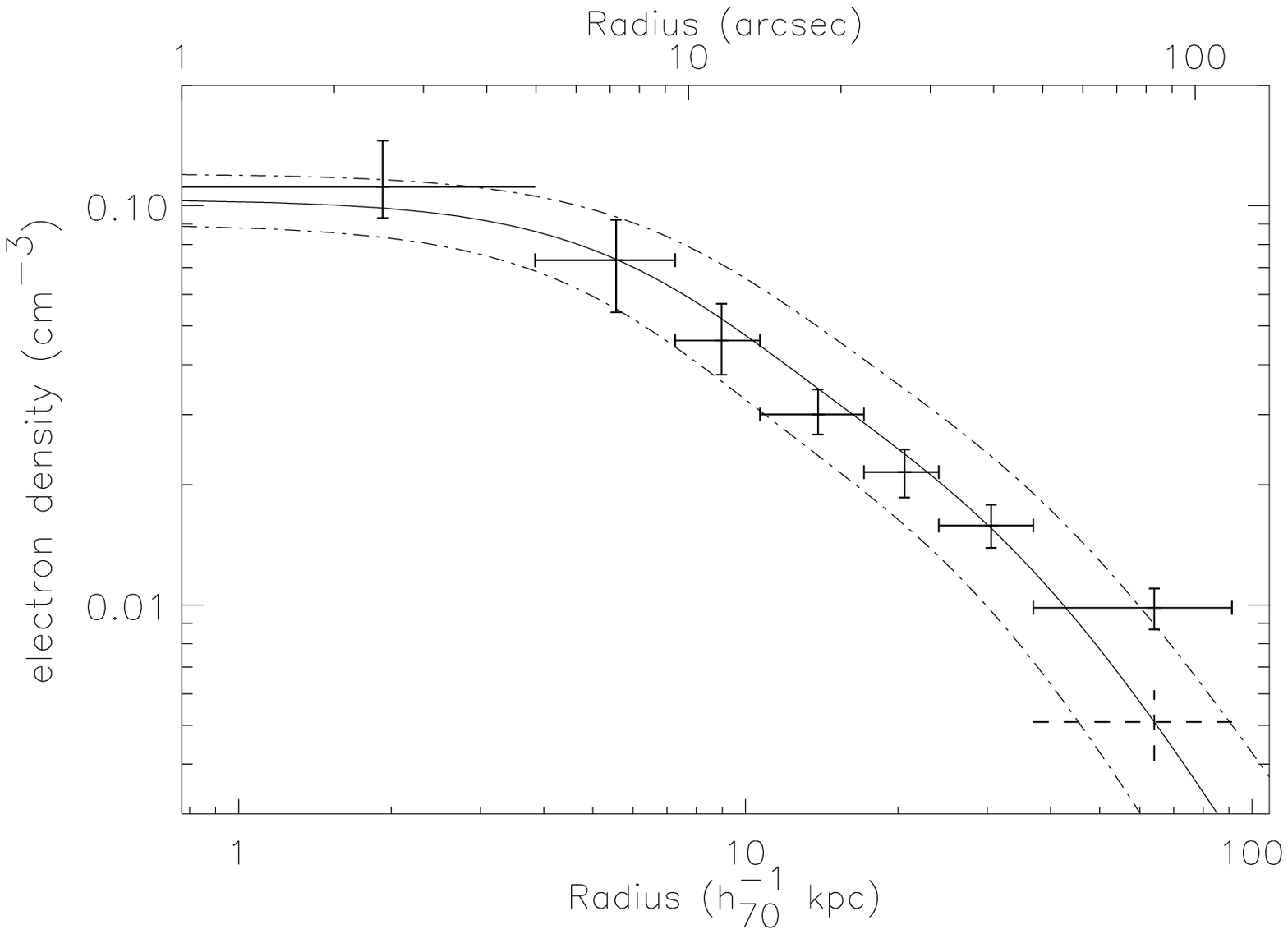}
    \includegraphics[scale=0.5,angle=0,keepaspectratio]{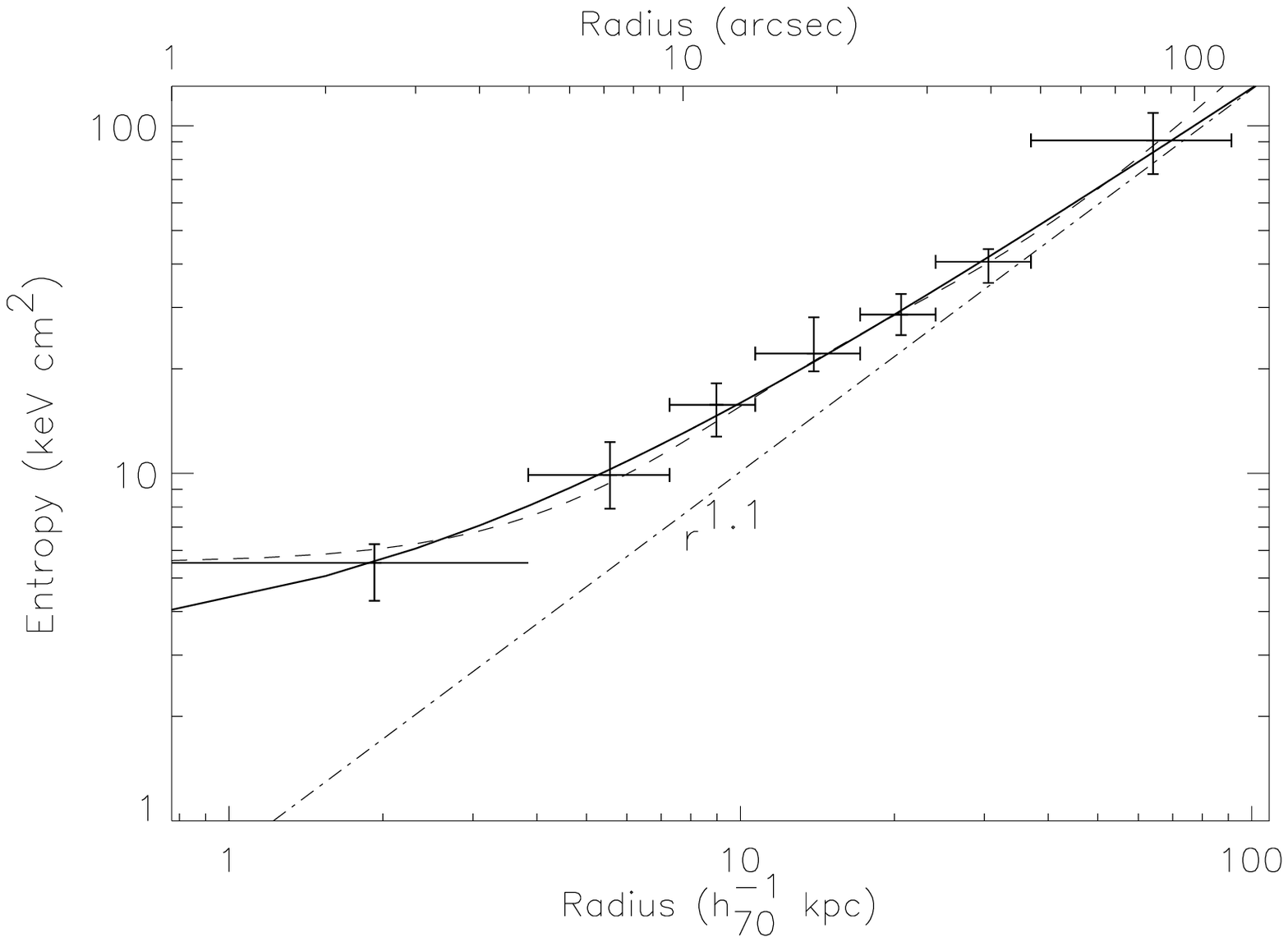}
    \includegraphics[scale=0.5,angle=0,keepaspectratio]{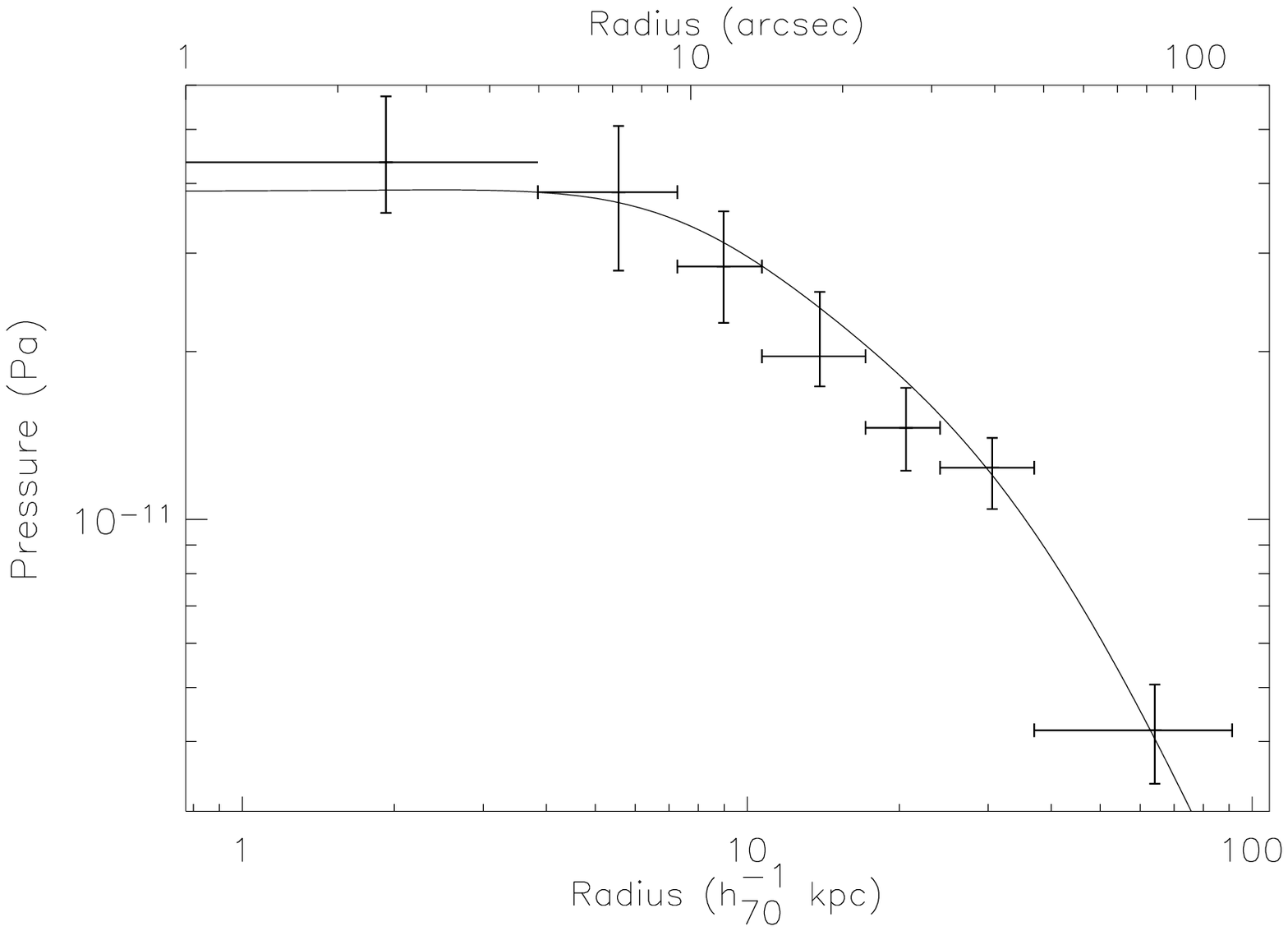}
\caption{ {\bf Top:} Electron density derived from the emission measure fitted in each annulus, assuming the density to be constant in each bin. The continuous line is the density profile derived from the double \betamod\ fitting of the surface brightness profile in the energy range 0.4-2.0 keV. Dot-dashed lines are the upper and lower 1$\sigma$ limits of this profile. {\bf Middle:} Entropy profile. The continuous line is the best-fitting power law (plus constant) model (see text). The dashed line represents the entropy profile derived by using the best-fitting temperature profile model and the \betamod-derived density profile. The dot-dashed line represents the entropy profile when the exponent is fixed to be 1.1 as expected from shock heating in spherical collapse (e.g. Tozzi \& Norman 2001; Voit et al. 2002); {\bf Bottom}: Pressure profile. The continuous line is the pressure profile derived from the deprojected temperature profile and  the density profile implied by the \betamod\  fit.}
\label{fig:nesp}
\end{centering}
\end{figure}

The derivation of the electron density directly from XSPEC is a suitable alternative option to the classical method of inverting the surface brightness profile into a density profile. The latter method requires us to estimate the emissivity function at each radius, and this is a strong function of the assumed temperature  for a cluster at $\sim 2.5$ keV, even if energies below 2.0 keV are used when generating the surface-brightness profile. We will see that the two methods are, in fact, in agreement.
In the top panel of Fig. \ref{fig:nesp} we show the electron density derived from the best-fit normalisation, assuming that the density is constant within each annulus. 
Since the  cluster extends well beyond our outermost annulus, we overestimate the electron density in the last bin. 
To correct for this effect we note that:

\begin{enumerate}
	\item the density derived from the isothermal \betamod\ described in Sec. \ref{sec:sbprof} is in agreement with the data points in Fig. \ref{fig:nesp}, for all annuli but the last (continuous line in Fig. \ref{fig:nesp}), despite the  clear temperature variations (Fig. \ref{fig:tzprof});
	
	\item the ROSAT data provide the best measures  of the cluster at larger scale, and the best-fitting $\beta$ found by Reiprich \& B\"ohringer (2002) is in excellent agreement with our $\beta$ value. This suggests that the extrapolation  of our best-fitting \betamod\ is a reasonable representation of the outer shells of the gas emission.

\end{enumerate}
For these reasons, we corrected the density in the last annulus to  equal  the \betamod-derived density at its distance from the centre (83 arcsec). We  added a 20\% systematic error to this bin.
The corrected density is 50\% lower than the original value, as shown in Fig. \ref{fig:nesp} by the dashed point. A similar argument applies to the proton density profile. We used the corrected values to calculate the entropy and pressure profiles below.

 We also note that the first data point is higher than (but consistent with) the \betamod\ profile (continuous line). This arises from the exclusion of the excess emission in the central 5 arcsec when fitting the $\beta$ model (second and third bin in  Fig. \ref{fig:sbprof}), while this excess was not excluded when we extracted the central spectral annulus (but note that the compact source and the extended structure $A$ described in Sec. \ref{sec:morphology} were masked).   

\subsection{Entropy and pressure profiles}\label{sec:PS}

The middle panel of Fig. \ref{fig:nesp} shows the profile of
$S = T~n_e^{-2/3}$ (Lloyd-Davies et al. 2000), which is related to the entropy per unit mass in the gas, derived from the temperature and density profiles obtained from the spectral fits. 

It is useful to fit the temperature profile with an analytical function. Here we use (Allen et al. 2001)

\begin{equation}
T = T_0 + T_1\times \frac{(r/r_c)^\eta} {(1+ (r/r_c)^\eta)}
\label{eq:tproffit}
\end{equation}
where our best-fit has $T_0=1.22\pm0.13$ keV, $T_1=1.40\pm0.17$ keV, and $r_c=(10.86\pm3.43)$ arcsec, close to the value of $r_c$ from the radial profile fit. The exponent, $\eta$=2, was fixed to the best-fitting value found by Allen et al. (2001). We then calculated the entropy profile using the best-fitting temperature model and the isothermal \betamod-derived electron density profile. This is shown in Fig. \ref{fig:nesp} by the dashed line. The dot-dashed line shows the S $\propto r^{1.1}$ behaviour found in  analytical and semi-analytical models (Tozzi \& Norman, 2001; Voit et al. 2002), and which is also found in numerical simulations without complex gas physics (e.g  Borgani et al. 2001). The normalisation is arbitrary.

A simple power-law fit of the form  $S= A \times r^{\alpha}$ to this entropy profile yields a slope $\alpha=0.77\pm0.04$ ($r$ is in arcsec). The effect of adding a constant ($S= A \times r^{\alpha}+ B$) is to increase the slope. The full details of the power law plus constant fit are $A = 1.04\pm0.27$, $\alpha = 0.99\pm0.07$ and $B = 3.02\pm0.73$. An $F$-test indicates that there is less than 1 per cent chance that the simple  power law is a better fit. 

It is now common to find the entropy profile scaled by $r_{200}$, the radius within which the mean cluster density is 200 times the mean density of the Universe. If we use the equation in Evrard, Metzler \& Navarro (1996), and the external temperature of the cluster (2.9 keV), we obtain $r_{200} = 1.4$ Mpc for \exo. This implies that our outermost annulus corresponds to $\sim0.05$ $r_{200}$, which is too small to allow useful comparisons with previous work on clusters with similar temperatures. Pratt \& Arnaud (2004) found $S\propto r^{0.94\pm0.14}$ for a small sample of clusters with $kT>2$ keV. Although this is calculated in the range (0.05-0.5) $r_{200}$, it looks consistent with the result we find here for \exo.

The bottom of Fig. \ref{fig:nesp} shows the pressure profile, where we compare the results when we use the XSPEC proton density profile (data points) and the isothermal \betamod\ density profile (continuous line).

\section{The central region}\label{sec:centre}
In previous sections we have described  the small-scale structure in the central 4 arcsec of \exo\ (see Fig. \ref{fig:imagauss}). In the surface-brightness profile (Sec. \ref{sec:sbprof}) we found excess emission above a \betamod, despite the fact that the compact source $P$ detected with {\sc wavdetect}, and the more elongated  emission labelled as $A$ in Fig. \ref{fig:imagauss}   were masked before extracting the radial profile. We also noted that the centre of the galaxy (and the cluster) displays a lower X-ray surface brightness than the immediate surroundings.
In this section we explore the nature of the compact source and the extended structure, and investigate the relation between the central gas and the radio source.

We extracted a spectrum of source $P$ from a circular region of radius 1.35 arcsec centred on  RA= $04^h25^m51^s02$, Dec = $-08^{\circ}33\arcmin36\farcs90$ (J2000). The local background was estimated in the surrounding annular region within  radii 1.35 and 2.95 arcsec. There are only 38 net counts in this region. Adopting C-statistics we found that a power law of spectral index 2.5$^{+1.3}_{-0.8}$ is a reasonable description of the data (the column density was fixed to the global value). With this model the unabsorbed 0.2-10 keV flux of the source is 3.9$\times10^{-14}$ ergs s$^{-1}$ cm$^{-2}$. A thermal model also describes the data well, but the temperature is not constrained.
With the data available, we can not state if the source is  a clump of thermal gas within the cluster or  a background (foreground) quasar, although the non-thermal interpretation seems more plausible from the spectral analysis. Given the positional offset between this source and the radio emission, it is unlikely that this is the BL Lac object we were looking for. 
We  mask this source for the following analysis.

We also extracted a spectrum of  structure $A$. We have even fewer photons but the spectrum shows a clear thermal signature, and  a  temperature consistent with the temperature that we found for nearby gas when  this structure was excluded (Sec. \ref{sec:tprof}). Since the surface-brightness profile displays excess emission between 1 and 4 arcsec from the centre, even after  excluding  structure $A$, we suggest that $A$ is of the same nature as the excess surface brightness, and its brightness may be due to projection effects or the particular geometry of the source.

\begin{figure}
\begin{centering}
   \includegraphics[scale=0.48,angle=0]{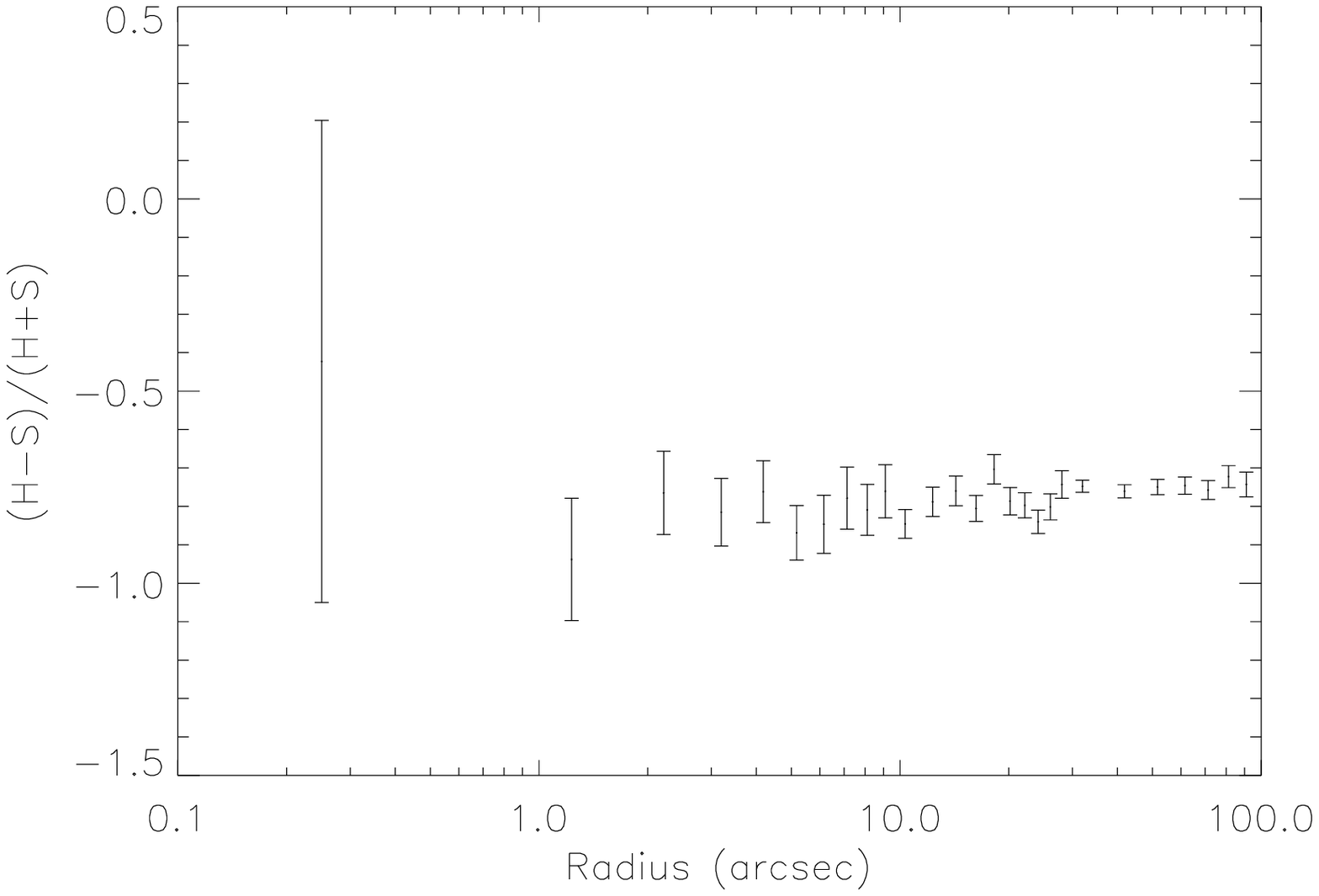}
    \includegraphics[scale=0.40,angle=0,keepaspectratio]{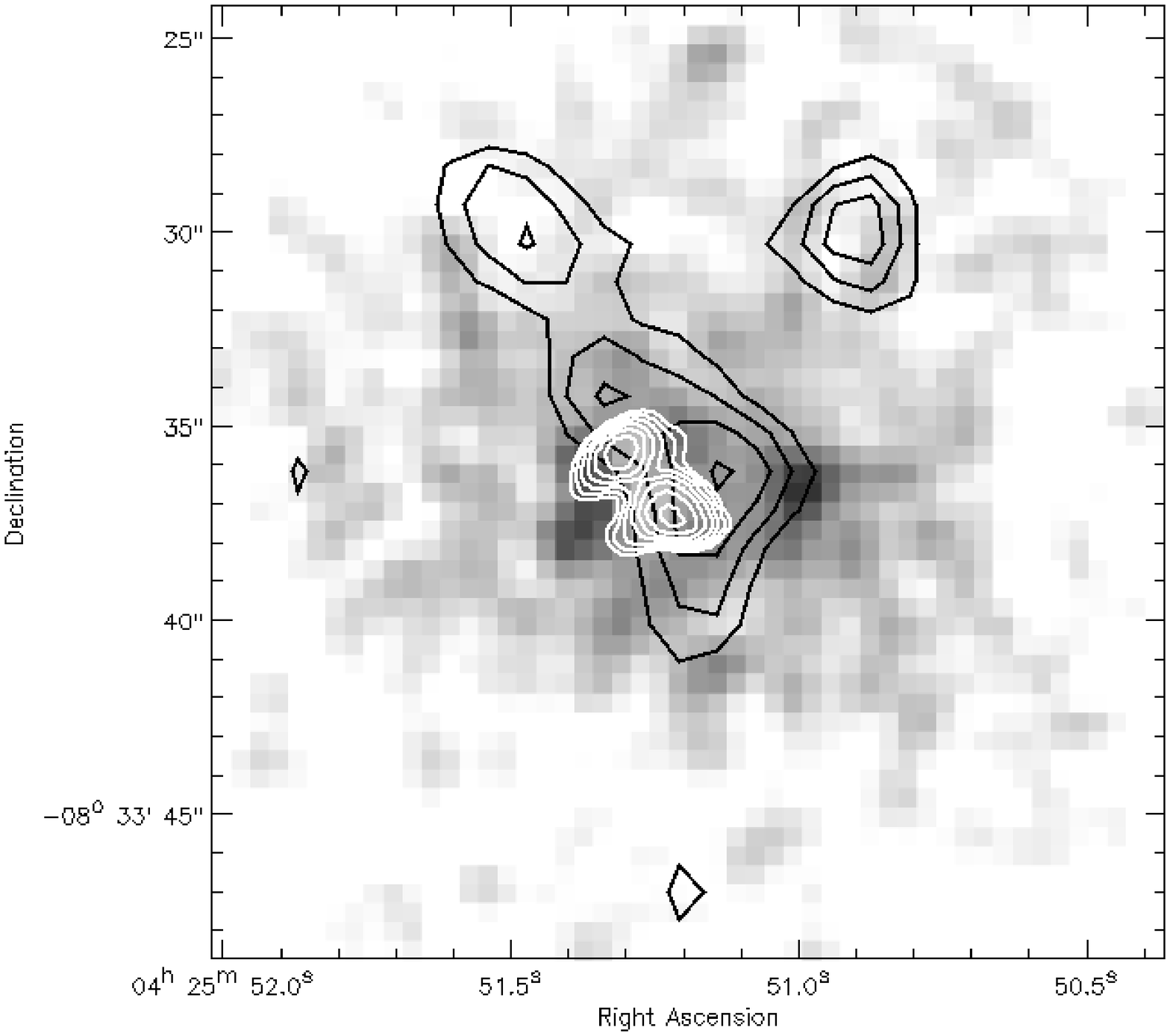}
\caption{ {\bf Top:} Hardness ratio profile. Radial bins are 0.5 arcsec wide out to 10 arcsec and then 5 arcsec wide. {\bf Bottom}: Image of \exo\ in the energy band 0.4-2.0 keV smoothed with a gauss filter of $\sigma=0.98$ arcsec. Black contours are obtained from the Gaussian filtered hard image with the same $\sigma$. White contours are from the 4.9 GHz map. The lower contour is traced at 0.5 mJy and they are logarithmically spaced up to 10 mJy. }
\label{fig:hr}
\end{centering}
\end{figure}

To investigate the spectral signature of the excess surface brightness, we extracted the surface brightness profile of \exo\ between 0 and 15 arcsec in smaller radial annuli than in Sec. \ref{sec:sbprof}, and in two energy bands: a soft band (S) at 0.4-2.0 keV,  and a hard band (H) at  2.5-6.5 keV. At larger scale the profile was binned in  5 arcsec wide annuli. We then obtained an hardness ratio (HR) profile by calculating  $(H-S)/(H+S)$ (Fig. \ref{fig:hr}). The average value of the HR is -0.77, reflecting the soft spectrum of the gas. Points at radius $R<5$ arcsec are consistent (within the statistical errors) with this average value. We generated S and H  images and smoothed them with a Gaussian filter of $\sigma$=2 pixels (0.98 arcsec).
In the bottom of  Fig. \ref{fig:hr} we show the soft image in grey scale. The black contours are obtained from the Gauss filtered hard image and the white contours are from the radio map at 4.9 GHz.  The hard emission is detected above the background at $4.5 \sigma$. We observe that the hard emission  is centred in the region of low surface brightness soft X-ray emission coincident with the optical centre of the galaxy, and that it is oriented in the same direction as the radio source but is more extended. The peaks of the hard X-ray and radio emission are not coincident, but this is likely to be a consequence of the smoothing, and is not significant. The peak to the north-west (black contours) is probably an unrelated source.

What is the physical relationship between  the radio source and the surrounding ICM?
First of all, there are two peaks in the 4.9 GHz radio map. We label as $S$ the southern component at RA = $04^h25^m51^s238$, Dec = $-08^{\circ}33\farcm37\farcs51$, and as $N$ the northern component at RA = $04^h25^m51^s325$, Dec = $-08^{\circ}33\arcmin35\farcs96$.  The  8.46 GHz map shows a peak of flux density 0.8 mJy between component $S$ and $N$, while components $S$ and $N$ are resolved and show steeper spectra. More sensitive radio observations would be needed to decide whether the core of the source lies in $S$ or $N$, or in the weak component that appears to lie between them. If we make the assumption that the source is in equipartition between radiating particles and magnetic field, we find an equipartition magnetic field of 7.4 nT and a pressure of 1.5$\times10^{-11}$ Pa for both components (we assumed a spectral slope of 2.2, minimum electron Lorentz factor of 100, maximum electron Lorentz factor of 10$^5$, and filling factor equal to 1). From the pressure profile discussed in Sec. \ref{sec:PS} the pressure of the external medium at the distance of each of the radio components from the optical centre is $\sim4\times10^{-11}$ Pa, which is only a factor of $\sim2$ from the internal minimum pressure of the source, suggesting that the source is close to pressure balance.

\section{Discussion}
We summarise below the results from our study of \exo\ using the high resolution of \Cha.

\begin{enumerate}
\item The X-ray emission from \exo\ is extended, without the strong point-like component expected for a BL Lac object. \exo\ is in fact associated with a cluster of galaxies, and an early type galaxy which dominates the central optical light;

\item we found absorption in excess of Galactic. This is not intrinsic to the source but of Galactic origin, as shown by the agreement between our X-ray measure and an independent measure which uses the \nh-IR emission relation;

\item we find that the  gas is locally isothermal, with a temperature decreasing towards the centre, as found for cooling-flow clusters, and with a temperature of $\sim3$ keV at $r>40$ kpc; however, a CF model gives a mass deposition rate of  only 10 $M_{\odot}$ yr$^{-1}$;

\item the metallicity profile declines outwards and it is consistent with an almost constant $Z=0.8Z_{\odot}$ at $r<30 h_{70}^{-1}$ kpc, and then drops  to 0.37$Z_{\odot}$; 

\item the gas entropy profile behaves as $S \propto r^{0.99\pm0.07}$, close to the relation predicted by analytical models of shock heating in spherical collapse, but shows a small core reflecting possible entropy modification in the centre;

\item the central radio source has a minimum pressure that is a factor of 2 below the external medium. Hard X-ray emission from the cluster centre has the same orientation as the radio source, but is more spatially extended.

\end{enumerate}

\subsection{ICM or IGM?}
The current \Cha\ observation gives us a precise look into the core of an intermediate temperature galaxy cluster.  The radial temperature profile is reminiscent of cooling core clusters, harbouring denser gas in the centre. Are we observing a classical cooling flow? Arguments against the classical cooling-flow models have been extensively discussed since the advent of \xmm\ (e.g. Tamura et al. 2001; Peterson et al. 2003; B\"ohringer et al. 2002). In our  case,  fitting a cooling-flow model to the global spectrum does not give a better fit than a two-phase model, but the temperature profile does suggest a cooling flow. 

Another interesting possibility is that we are observing the interstellar medium (ISM) of the early type galaxy. Indeed our  \Cha\ observation maps  X-rays only within  the optical envelope of the galaxy. A clear indication in favour of this interpretation is given by the metallicity profile. We are not able to map individual  elements  with the current data, but  the fitted metallicity should be dominated by the iron abundance. The \exo\ metallicity profile clearly shows a decline from the centre outwards, which can also be seen as a  break at $\sim 35 h_{70}^{-1}$ kpc, with an average global relative abundance of $\sim 0.8$ within this radius and $Z/Z_{\odot}$=0.37 outside. Work based on  ASCA data found that clusters harbouring cD galaxies display a significant increase in metal abundance towards the centre (e.g. Fukuzawa et al. 2000; Matsushita et al. 1998). Similar results were found with BeppoSax data (e.g. De Grandi  \& Molendi 2001), and suggest that this is due to metal enrichment by SN type Ia in the cDs (e.g. Finoguenov \& Ponman 1999, Matsushita et al. 2002, 2003).
The optical galaxy that we associate with \exo, MCG-01-12-005, is not well studied and it is not catalogued as a cD galaxy. However, this interpretation is still plausible, since prior to the advent of \Cha\  it was almost impossible to map in such detail the inner regions of galaxy clusters, and  objects without the striking optical properties of cD galaxies  are just starting to be discussed. 

An acceptable estimate of the galaxy X-ray temperature is given by the first bin of the temperature profile (k$T$$\sim$ 1.3 keV). We find a corresponding X-ray luminosity of $\sim 3.6\times10^{42}$ erg s$^{-1}$, once projection is taken into account, and after  extrapolating to 40 kpc (when the cluster emission dominates). Although the temperature is rather high, the values we find are in agreement with the $L_X-T$ relation of O'Sullivan et al. (2003) for early type galaxies, adding another piece of support to the ISM interpretation for the X-ray emission within 40 kpc of this object. While the temperature increases rapidly after the first bin, and at 40 kpc it is already $\sim2.5$ keV which is too high for the atmosphere of a galaxy, we note that the metallicity and temperature profiles could change independently with different radial functions because of the different physics associated with the transport of heavy ions outwards and thermal energy inwards.

We also note that the  gas is consistent with a two-phase metallicity, with a value close to solar within 35 arcsec and close to a common value for galaxy clusters beyond 35 arcsec. This suggests that little mixing between the ISM and ICM has occurred.
 
\subsection{The impact of the radio source}

We can use the comparison between the radio source  and  the ambient medium properties  to map the formation history of the core of \exo.
We find that the gas entropy declines continuously towards the centre, in agreement with recent results on galaxy groups and clusters (e.g. Ponman, Sanderson \& Finoguenov 2003; Pratt \& Arnaud 2003; Khosroshahi, Jones \& Ponman 2004, Pratt \& Arnaud 2004), typically of the same temperature as \exo. Moreover, the large-scale radial dependence of the entropy we measure is in agreement with both analytical models of spherical collapse, and numerical simulations based solely on gravitational collapse (Tozzi \& Norman, 2001, Borgani et al. 2001, Voit et al. 2002). It does seem that there is or there has been some modification of the core entropy of this cluster (galaxy). This is seen in the radial dependence of the entropy, where a simple power-law fit gives a shallower slope than expected from pure shock heating, and where a power law plus constant model is clearly a better fit. However there is  no flat  isentropic core (entropy plateau).

The deviation of other classical scaling laws, such as the $L_{\rm X}-T$ relation (e.g. Arnaud \& Evrard 1999), from the prediction of purely gravitational models has suggested that some extra heating has occurred. 
It has been proposed that sources at the centres of clusters (such as AGN or radio galaxies) could cause this heating (e.g. Valageas \& Silk 1999; Wu, Fabian \& Nulsen 2000; Churazov et al. 2002; Ruszkowski \&  Begelman 2002).  The  small core in the entropy distribution of this object would suggest that entropy has been modified in some way, and the central radio source, although weak, may be a plausible candidate for this modification (e.g. Roychowdhury et al. 2004).

An alternative observational indication is given by the comparison between the radio and X-ray luminosity in the cooling region, which should be similar, once a reasonable factor for the conversion between power and radio emission is assumed. The radio source at the centre of \exo\ is particularly small.  The radio luminosity is of order $5\times10^{39}$ erg s$^{-1}$, while the X-ray luminosity of the soft thermal component  within 35 arcsec is $\sim 3.6\times10^{42}$ erg s$^{-1}$. If only 0.1\% of the total  power output goes into the radio emission, then the rest of the power can heat the gas. However, this is a rather low efficiency for conversion of the source power into radio emission, and also a high efficiency for heating. From this evidence, it is unlikely that the  radio source is supplying sufficient heat to stop any cooling flow.

\subsection{The age of the radio source}
Since we do not have an independent measure of the magnetic field,  we assumed equipartition, yielding  the result that the minimum internal pressure of the source is about a factor of 2 lower that the external pressure. 
The minimum pressure is a lower limit, thus  we are facing two possibilities for the small size of the source: the radio source is old and is dying, or the radio source is young.

In the first case, the source was bigger and more powerful in the past. It may have heated the external medium which is now hotter than at the origin, as can be inferred from the relative distribution of  hard and soft emission in the core of \exo\ (Fig. \ref{fig:hr}). It is also possible that  the source inflated bubbles of rising gas, carrying energy and pressure, as observed for example in M87 (e.g Ghizzardi et al. 2004). However, radio sources are short-lived (10$^7- 10^8$ yr) and when they are starved of fuel they lose power. 
If this is the scenario we are observing, the radio source should be relatively old, and is in its fading phase. Hints in favour of this picture are the small size of the radio source, the  distribution of the hard emission in the central 5 arcsec (Fig. \ref{fig:hr}), the small entropy core, and the relatively large external pressure at the location of the radio galaxy. We cannot constrain the spatial distribution of the hard emission, which is strongly limited by the statistics. It is unlikely that the hard emission is non-thermal, since high energy electrons  cannot live longer than their parent population. Thus it is possible that what we are observing is the signature of gas heating. 

The second possibility is that the source is  young. The engine has just turned on and the source is expanding. This would explain i) the low  surface brightness of the gas at the location of the radio source, and ii) the high surface brightness just outside it. This can be interpreted as an expanding source which is pushing the ambient gas away.
If the source is young then it may not have had time to affect the entropy of the gas, which could explain the relatively small  heating signatures in the entropy profile.

\section{Conclusions}
The \Cha\ observation of \exo\ shows that a BL Lac object is absent and we observe an extended, cluster-scale source, which was previously discussed by Edge \& Stewart (1991). The central X-ray emission is likely to be due to the ISM of the early-type galaxy at the centre of the cluster, and only at $r>40$ kpc does the ICM dominate over the ISM. The central X-ray emission was little affected by heating in the way observed for other low-temperature clusters. If the source is relatively young, our results represent  indirect support for models where AGN activity in the central galaxy may be cause of  the breakdown of self-similarity.

\section*{Acknowledgements}
We are very grateful to G.W. Pratt for many helpful discussions. We thank M.J. Hardcastle for helping in the initial phase of the work. We thank an anonymous referee for a thorough referee's report which enabled us to improve the manuscript. E.B. and D.M.W. thank  PPARC for support. This research has made use of the SIMBAD database, operated at CDS, Strasbourg, France, and the NASA/IPAC Extragalactic Database (NED) which is operated by the Jet Propulsion Laboratory, California Institute of Technology, under contract with the National Aeronautics and Space Administration. The National Radio Astronomy Observatory is a facility of the National Science Foundation operated under cooperative agreement by Associated Universities, Inc.

\clearpage
\end{document}